\documentstyle[pre,aps,floats,graphics,epsfig]{revtex}
\newcommand{\be}{\begin{equation}}
\newcommand{\ee}{\end{equation}}

\title{Self-referential behaviour, overreaction and conventions in financial markets}
\author{Matthieu Wyart$^1$, Jean-Philippe Bouchaud$^{1,2}$}

\address{
 $^{1}$ Service de Physique de l'\'Etat Condens\'e\\
Orme des Merisiers --- CEA Saclay, 91191 Gif sur Yvette Cedex, France.\\
}
\address{
 $^{2}$ Science \& Finance, Capital Fund Management\\
109--111 rue Victor-Hugo, 92532 Levallois Cedex, France\\
}
\date{\today}
\draft

\begin{document}
\maketitle

\begin{abstract}
We study a generic model for self-referential behaviour in financial markets, 
where agents attempt to use some (possibly fictitious) causal correlations between 
a certain quantitative information and the price itself. This correlation
is estimated using the past history itself, and is used by a fraction of
agents to devise active trading strategies. The impact of these strategies on the
price modify the observed correlations. A potentially unstable feedback loop appears and destabilizes 
the market from an efficient behaviour. For large enough feedbacks, 
we find a `phase transition' beyond which non trivial correlations spontaneously set in and where the market 
switches between two long lived states, that we call conventions. 
This mechanism leads to overreaction and excess volatility, which may be considerable
in the convention phase. A particularly relevant case is when the source of information is the price itself. 
The two conventions then correspond then to either a trend following regime or to a contrarian
(mean reverting) regime. We provide some empirical evidence for the existence long lasting 
anomalous correlations in real markets, which reflect the existence of these conventions.
\end{abstract}

\section{Introduction}
\label{sect:intro}

\subsection{Aim of the paper} 

Efficient Market Theory claims that prices contain all available information at a given 
instant of time (for more precise statements, see \cite{Fama}). The argument invoked to support 
this claim is arbitrage: if prices 
differed from their informationally efficient value, an arbitrageur possessing some information
not reflected in the price would be able to make a profit, and doing so would bring the 
price closer to its true value. In this framework, price changes can only be triggered 
by some new, unpredictable piece of information. Therefore, as shown by Samuelson \cite{Samuelson}, 
{\it properly anticipated prices fluctuate randomly}, i.e. prices should follow a random 
walk. This theory relies on the assumption that all agents are rational and are all seeking to
discover the `true', fundamental value of a stock.

However, this assumption is extremely strong and has been criticized by many authors (see,
e.g. \cite{Schiller,Shleifer}). It seems clear that many agents in fact do not behave like this. One reason is 
that the number of objective factors that can affect the value of a stock is very large, 
and that 
interpretation to be given to some `information' is often totally ambiguous. Therefore, as
emphasized by Keynes and more recently Orl\'ean \cite{Keynes,Orlean}, market participants are more 
interested to
guess the opinion of the market than to discover the fundamental value of the stock. As 
illustrated by Keynes' famous beauty contest \cite{Keynes}, the goal is to correctly anticipate what
other participants do anticipate; this self-referential behaviour can lead to markets that 
differ strongly from the predictions of Efficient Market Theory. 
An interesting example is provided by a simple game that encapsulates the basic 
message of Keynes' beauty concept. In this game, participants must
each choose a number between $0$ and $100$, and the winner(s) are those whose choice is closest to 
one-half of the average choice \cite{Nagel}. Of course, the fully rational choice is that all players 
choose $0$. On the other hand, if agents are all totally irrational, the optimal choice is $25=50/2$, 
but if a fraction $f_1$ follows this first level reasoning, the optimal choice becomes 
$[25f_1+50(1-f_1)]/2$, etc. Empirical studies show that the average is close to $25$, and that
$30\%$ of the players predict a number close to $12.5$. In the context of 
repeated games (such as financial markets), a natural strategy 
is to study empirically the statistical behaviour of the other agents and 
to play accordingly. (Agents doing this will be called `strategic' in the following). A 
common temptation is to compare the present situation
with similar situations from the past, and make the guess that what already happened is more
likely to happen again. As Brian Arthur puts it: {\it As the situation is replayed regularly, 
we look for patterns, and we use these to construct temporary expectational models or 
hypotheses to work with} \cite{Arthur}. 
For example, price tend to decline before a war is declared, and to
rise again once the war has actually started (as recent events again sadly confirm). 
Often, some plausible story is given to understand 
why such a pattern should exist. This convinces more participants 
that the effect is real and their
resulting behaviour is such to reinforce (or even to create) the effect: this is a 
self-fulfilling prophecy. A large consensus among 
economic agents about the {\it correlations} between a piece of information and the 
market reaction 
can be enough to establish these correlations. Such a `condensation' of opinions leads to what Keynes 
and Orl\'ean 
called a {\it convention} \cite{Keynes,Orlean}, a common lore on which uncertain
agents can rely on, and that supplements gossamer information. A convention may 
concern the overall mood of the market (bullish or bearish, for example), but may also
concern the way a piece of information is interpreted by the market. We will primarily 
focus on this second type of convention. The information we will consider can either exogenous 
to the market (such as the interest rate, 
inflation and other macro-economic figures, or geopolitical issues), or 
endogenous to the market, such as price patterns that feedback on the price itself (trends leading to bubbles,
or ARCH-like volatility feedback, etc.) 

A striking feature is that not only these conventions spontaneously appear, but can
also disappear or even invert the purported correlation. For example, as we document in Section V below
(see in particular Fig. 5), the correlation between bond markets and stock markets was positive in the past (because low 
long term interest rates should favor stocks), but has recently quite suddenly become negative as
a new `Flight To Quality' convention set in: selling risky stocks and buying safe bonds 
has recently been the dominant pattern.

The aim of the present paper is to analyse a parsimonious model for the appearance and 
dynamics of 
conventions, and their consequence on the statistics of price changes. As is now well known, 
price returns exhibit several statistical features that cannot be related to the fluctuations of 
any fundamental value, as should be if markets were efficient \cite{Fama}. One of the biggest puzzle 
of the efficient market theory is the so-called `excess volatility': Schiller 
showed in a famous study that the actual volatility of markets is far too large 
compared to what is expected, within the efficient market theory, from the volatility of dividends \cite{Schiller1}. Even accounting 
for some reasonable uncertainty about the expected value of these dividends leaves the 
empirical volatility at least a factor $\approx 5$ too large \cite{Schiller}. Also, the volatility is itself random 
and fluctuates in time, exhibiting `clustering' and remarkable long-range temporal 
autocorrelations (long term memory) \cite{volfluct1,volfluct2,Guillaume,Cizeau,PCB,Rama,Book,LeBaron}, analogous to fluctuations in turbulent flows \cite{Mandel3,BMD,Lux2}. Other similar effects are
worth mentioning, such as the excess cross-correlations, both between domestic
stocks and between international markets, that cannot be explained in terms of fundamental,
economic correlations \cite{Schiller,Ang,Longin}. 
 
Our model is an example of a self-fulfilling process: trying to extract correlations between 
information and price from past observations, market participants tend to create and/or reinforce them. 
Using the language of physics, the model has a {\it phase transition}: 
above a certain threshold in 
feedback strength, the market can be in two distinct states, or two conventions. We find that 
this mechanism 
naturally leads to some excess volatility and long term memory. In the case where 
information is 
endogenous, these two states correspond to trend following or contrarian `conventions', where the
autocorrelations are either positive (trend) or negative (contrarian). The market however 
switches 
between the two conventions on a certain time scale (that can be very long) such that 
{\it on average} 
the autocorrelation is zero, although {\it locally} the autocorrelation has a well defined sign. 
These phases correspond to the market folklore: markets are indeed thought by many investors 
to be alternatively `trending' or `mean-reverting', in strong opposition with the prediction 
of Efficient Market Theory. Beside anecdotal evidence, 
we provide in this paper convincing empirical evidence for the existence of these long-lived excess 
correlations (or anticorrelations) in stock markets (see Section V, Figs 6,7).

\subsection{Relation with other work and organization of the paper}

The existence of trends and `anti-trends' has been broadly documented in the economic
literature (see e.g. \cite{Thaler,Shleifer,Stein} and refs. therein), where it is described as 
`overreaction' and `underreaction' to news. In the first case, the overreaction is later on 
compensated by a mean reversion, whereas underreaction corresponds to an anomalously slow 
adjustment of the price and the appearance of a trend. A well known study of de Bondt and Thaler 
\cite{Thaler}, that we will further discuss in the conclusion,
shows that over-performing stocks tend to `mean-revert' on the scale of 5 years, and vice-versa.
Several `behavioral' models have
recently been proposed to understand these effects (see \cite{Shleifer}, chapters 5 \& 6, and
\cite{Shleiferpaper,Stein}). We follow here the same goal of articulating a simple and generic model 
for these pricing anomalies. Although some ingredients are common to these models and ours, there are
also major differences, both in the concepts and in their technical implementation. For example, in the `investor sentiment' model of \cite{Shleiferpaper}, 
investors postulate the existence of alternating `trending' and `mean-reverting' phases that 
they try to identify from observation. In our framework, on the other hand, these phases 
dynamically appear as agents attempt to learn the statistics of price changes from past
observations. This `learning' aspect, much emphasized in \cite{Arthur,MG} is actually 
absent from the model presented in \cite{Shleiferpaper}. The model of Hong and Stein \cite{Stein}
postulates that some `momentum traders' use (but only for a limited amount of time) the positive temporal 
correlations created by the
slow diffusion of information among `news-watchers'. The effect of these momentum traders 
is to reinforce the trend and to convert an initial `underreaction' into `overreaction'. 
In order to observe mean reversion effects, another category of `contrarian' traders 
must be put by hand. In our model, on the other hand, `trends' can appear without any fundamental news. 
In this respect, it is perhaps useful to mention the work on information cascades \cite{cascades}, 
which, although very different in spirit, also describes a situation where a symmetry between
two possible outcome can be broken by a small initial bias, amplified by a subsequent
self-referential decision process. Finally, several 
families of models where trading strategies use the price past history have recently 
been investigated, for example in the schematic inductive rationality models (El Farol bar model
\cite{Arthur} or Minority Games \cite{MG}) or in agent based models where a fraction of 
agents base their trading decision on the recent behaviour of the price 
itself \cite{SantaFe,Langevin,Farmer2,Stein,Lux,Hommes,Chiarella,Iori,GB}. The present model is interesting 
because the self-referential feedback is much simpler and its consequences can be analytically
investigated in full details.

The organization of the paper is as follows. We describe and analyze the model in full 
generality in Section II,
when agents try to use some correlations between the price and a certain information 
indicator, which can 
be exogenous or endogenous. We motivate our `Langevin' description of the feedback dynamics and explain how non trivial `equilibria' can appear when the self-referential tendency increases. We discuss the appearance of super-long time scales for regime switching (Section III). 
We then specialize to the particular case where 
traders use the 
past price changes as a source of information (Section IV), and where the above mentioned trend 
following or contrarian `conventions' (or market sentiments) appear. In some parameter region, 
trends can be short in time but strong in intensity, which leads to large price jumps, or crashes. We then analyze some empirical data that support the predictions
of the model (Section V). Finally, some extensions of the model are proposed, and our findings are 
contrasted with the predictions of Efficient Market Theory.

\section{Set up of the model}
\label{sect:model}

We will call $P_t$ the (log-)price of a certain asset at time $t$, and $\delta P_t$ the  
return between $t$ and $t+1$. Here, $\Delta t=1$ is the elementary time step over which 
agents revise their strategies, which might 
be one day or one week, although in some case smaller time scales (like minutes) could also
be usefully considered. We now argue that some agents base their strategy on the observation
of the temporal change of a certain `index' $I_t$, which might be a financial index or 
an economic indicator (for example dividends, interest rates, 
inflation, confidence, unemployment, etc.), or even, as
will be considered below, the price $P_t$ itself. We will denote as $\delta I_t$ the change of this 
indicator between $t$ and $t+1$. Note that $\delta I_t$ could in fact be a binary variable, representing
a qualitative piece of information, and that the interval $\Delta t$ might not be uniform, and
be the time interval between the arrival of news. 

Suppose that there exists a causal correlation between the change of $I_t$ and that of
$P_t$, in the sense that the correlation between $\delta I_t$ and $\delta P_{t+1}$:
\be
E\left[\delta I_t \delta P_{t+1}\right] \equiv C. 
\ee
(We suppose for simplicity that all correlations for larger time lags are zero).
It is then a well known result of linear filtering that the best estimate (in a quadratic 
sense) of $\delta P_{t+1}$ knowing $\delta I_t$ is given by (see e.g. \cite{Book}, p. 132):
\be
\delta P_{t+1}^* = \frac{C}{E\left[\delta I_t^2\right]} \, \delta I_t. 
\ee

Now, we consider two types of agents, those who act randomly, or based on 
some information uncorrelated with $I_t$, and those 
who try to take advantage of the possible correlations between $\delta I_t$ and 
$\delta P_{t+1}$.
Since the `fundamental' value of the correlation $C$ is in fact not known, agents 
of the second type attempt to extract this value from past history, from which they 
try to learn the value of $C$. It is natural to
assume that these agents give more weight to the recent past. A convenient framework
is that of exponential moving averages, such that the {\it estimated} value of $C$ 
at time $t$ is given by:
\be\label{Ct}
C_t = \frac{1-\alpha}{\alpha} 
\sum_{t'=-\infty}^{t-1} \alpha^{t-t'} \delta I_{t'} \delta P_{t'+1},
\ee
where $\alpha$ sets the memory time $T$ of the agents, as $T=1/|\ln \alpha|$. 
Eq. (\ref{Ct}) is equivalent to the following Markovian update of the estimated 
correlation:
\be
C_t = \alpha C_{t-1} + (1-\alpha) \delta I_{t-1} \delta P_{t}.
\ee
We now suppose that the agents neglect the possible fluctuations of the volatility 
of $I_t$ and assume $E\left[\delta I_t^2\right]$ is a constant (that we set in the following 
equal to unity, unless stated otherwise). Relaxing this hypothesis 
would lead to minor changes in the following. The expected return between
$t$ and $t+1$ is therefore $C_t \delta I_t$; it is natural to assume that one will
buy (or sell) a quantity $V_t$ which is an odd function of this expected return:
\be
V_t = {\cal G}\left(C_t \delta I_t\right).
\ee
In general one expects the demand function ${\cal G}$ to be linear for small arguments and to saturate 
for large arguments. In the context of an exponential utility function (called CARA in the literature), 
the quantity to be maximized is the expected return minus a certain coefficient times the 
variance of the return. In this case, the function ${\cal G}$ is found to be strictly linear. 
The saturation comes from both the limited resources of the agents and their limited ability to borrow and from an increased risk aversion for tail events \cite{Potters}. Both effects tend to limit the invested quantity even if the signal is very strong. These strategic orders add to 
the non strategic ones and impact the price as:
\be
\delta P_{t+1} = {\cal F}\left(\Omega_t+ N V_t\right),
\ee
where $N$ is the number of `strategic' agents that try exploit this correlation, 
and $\Omega_t$ the total volume of
`non strategic' agents, which we assume to be a random variable of zero mean 
and variance $\sigma^2$. (In fact, as we will discuss below, these non strategic
agents could base their decision on other, uncorrelated information sources). 
The {\it impact function} ${\cal F}$ describes how a given 
order volume affects the price, and has been the subject of many recent empirical
studies \cite{Kempf,Gopi,Lillo,Bali}. Provided the elementary time step $\Delta t$ is 
large enough, 
this function is linear for small arguments and bends down for larger order imbalance.
Here, we will neglect higher order contributions to ${\cal F}$ and simply posit, as in 
\cite{BG,Langevin,Farmer2}:
\be
{\cal F}(u)=\frac{u}{\lambda},
\ee
where $\lambda$ is a measure of the liquidity of the asset. Higher order corrections would only change 
some details of the following discussion. Since ${\cal G}$ is odd, its generic expansion for
small arguments reads ${\cal G}(u)=a u - b u^3 + ...$ with $a,b >0$, with higher order terms that 
do not change the following qualitative conclusions. We finally obtain the central equation of the present study, valid in the small signal limit:
\be\label{price}
\delta P_{t+1} =\frac{\Omega_t}{\lambda}+ g C_t \delta I_t - h C_t^3 \delta I_t^3 + O(C^5),
\ee
where $C_t$ is self-consistently expressed as (\ref{Ct}), and $g \equiv Na/\lambda$, $h 
\equiv Nb/\lambda$.
These two equations basically describe the self fulfilling process that we study in details now.
The parameter $g$ will turn out to be crucial in the
following; note that $g$ increases with the number of strategic agents.

\section{analytic results: spontaneous appearance of conventions} 
\label{sect:results}

\subsection{A Langevin equation}

In the absence of strategic agents ($g=0$), there are no feedback effects, and the
dynamics of the price is a simple random walk of volatility $\Sigma_0=\sigma/\lambda$. The 
apparent correlation $C_t$ will describe any deviation from this trivial behaviour. 
Using Eqs. (\ref{Ct},\ref{price}) one finds:
\be
C_{t+1} - C_{t} = \epsilon \left(-C_t + g C_t \delta I_t^2 - h C_t^3 \delta I_t^4 + \xi_t\right),
\ee
where we have set $1-\alpha = \epsilon$, and $\xi_t \equiv \delta I_t \Omega_t/\lambda$ is another white noise 
(because $\Omega_t$ is supposed to be independent of $\delta I_t$), of zero mean and variance $\sigma^2/\lambda^2$.

Now, we will write:
\be
\delta I_t^2 = E(\delta I_t^2) + \eta_t = 1 + \eta_t; \qquad 
\delta I_t^4 = E(\delta I_t^4) + \eta'_t = (3+\kappa) + \eta_t',
\ee
where $\eta_t$ and $\eta_t'$ are two correlated noises of zero mean, and $\kappa$ the
excess kurtosis of the index fluctuations. Therefore the evolution of $C_t$ contains 
a deterministic part and a random part. In the case $\epsilon \ll 1$ considered in this paper, 
where the memory time $T \approx 1/\epsilon$ becomes much larger than the elementary 
time step, one can neglect, in a first approximation, the influence of $\eta_t$ and $\eta_t'$ 
(but see below). Taking the continuous time limit $\epsilon \to 0$, one can write a Langevin stochastic differential equation for $\hat C_t=C_t/\Sigma_0$ in rescaled time $\epsilon t = \hat t$:
\be
{d\hat C} = - \frac{dV}{d\hat C} {d\hat t} + \sqrt{\epsilon} d \xi,
\ee
where $d \xi$ is a Brownian noise of unit variance and the `potential' $V$ is given by:
\be
V(\hat C) = \frac{1}{2} (1-g) \hat C^2 + \frac{1}{4} (3+\kappa) \hat h 
\hat C^4, 
\ee
with $\hat h \equiv h \Sigma_0^2$. This is the so-called Landau potential that describes phase transitions \cite{Goldenfeld}. For $g < 1$
this potential has an absolute minimum at $\hat C=0$, whereas for $g > 1$, $\hat C=0$ becomes 
a local maximum and two stable minima appear for $\hat C=\pm C^*=\pm \sqrt{(g-1)/(3+\kappa)\hat h}$. Note that retaining more terms in the expansion of ${\cal G}$
would change the detailed shape of $V(\hat C)$, but not the above crucial qualitative feature.
From now on, we will drop the hat on $\hat C$.

\subsection{The appearance of stable conventions}

From the Langevin equation for $C$ one deduces, using standard methods \cite{Chandrasekar,Gardiner}, the equilibrium distribution $P(C)$ which 
is of the Boltzmann-Gibbs form:
\be\label{Bolz}
P(C) = \frac{1}{Z} \exp \left(-\frac{2 V(C)}{\epsilon}\right),
\ee
where $Z$ is a suitable normalization. Therefore, for $g < 1$ (weak feedback), 
$P(C)$ is unimodal and has a maximum at $C=0$, whereas for $g > 1$ (strong feedback), 
the most probable values for $ C$
are $\pm C^*$. This means that for strong feedback, a non zero correlation between the price 
and the indicator spontaneously appears. This correlation can be either positive or negative, 
corresponding to the two 
possible `conventions'. However, on average, the
correlation is still zero for $g > 1$, since $C$ randomly flips between $\pm C^*$. 
In order to do so, 
a `potential barrier' $\Delta V$ has to be crossed (see Fig. 1); the `switching' time $\tau$ is well known to be
given, for $T\Delta V \gg 1$, by the Arrhenius law \cite{Chandrasekar,Gardiner}:
\be\label{Arrh}
\tau \simeq \left(\frac{T}{g-1}\right) \, \exp \left[{2 T \Delta V}\right], 
\ee
with $\Delta V = (g-1)^2/4\hat h(3+\kappa)$ and $T=1/\epsilon$. Because of the exponential term, this switching time
can be much larger than the memory time $T$:
one non trivial consequence of a phase transition is to generate time scales that are unrelated
to the natural time scale of the problem. The convention can therefore persist for 
very 
long times.
This is because the random event that would `invert' the signal and nucleate a new 
convention occurs
only exponentially rarely. Note however that the above formula is only correct when 
the noise 
$\xi(t)$ is Gaussian; non Gaussian events do accelerate the crossing of the barrier \cite{Anusha}. 
We will see in empirical 
data that extreme events may indeed be a cause of abrupt convention changes.

\begin{figure}
\begin{center}
\psfig{file=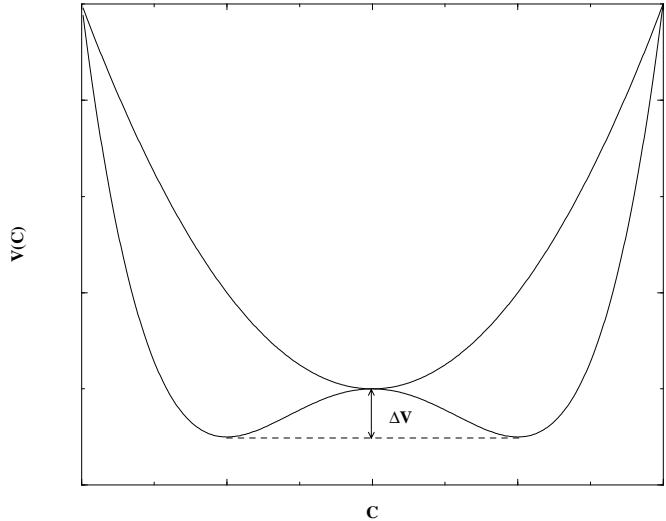,width=7cm,angle=270} 
\end{center}
\caption{Effective `potential' $V(C)$ for $g < 1$ and for $g > 1$. In the latter case, one observes two non trivial minima at $\pm C^*$ and a `potential barrier' $\Delta V$ separating them.}
\label{Fig1}
\end{figure}

When $g < 1$, the distribution of $C$ is a distorted Gaussian around $C=0$. Neglecting the 
non linear term leads to:
\be
C_t = \epsilon \, \int_0^t {\mbox d}\xi(t') \,
{\mbox e}^{\epsilon (g-1)(t-t')},
\ee
where we have assumed for simplicity $C_{t=0}=0$. Hence the typical value of $C$ is $ C \sim \sqrt{{\epsilon}/{(1-g)}}$ and typical time 
$\tau={T}/{(1-g)}$ for the variations of $C$, which diverges when $g$ gets close to 1. 
The strategic agents thus amplify the excursions of $C$, but the most probable value of $C$ 
is still zero. Strictly speaking, there is no stable point or convention in this case, although
when $g \to 1^-$ the excursions are of larger amplitude and of longer duration, which 
corresponds to what can be coined a `floating convention'.\footnote{We should add here a remark on the r\^ole of the multiplicative noise term $g C_t \eta_t$ 
neglected in the above analysis. Since $\Omega_t$ and $\eta_t$ are uncorrelated, the Langevin 
noise has a variance now given by $\epsilon[1+ (2+\kappa) g^2 C^2]$. Going to the
Fokker-Planck equation \cite{Chandrasekar}, one can show that for small $\epsilon$, the r\^ole of this extra term is
to shift the value of the critical threshold to $g = 1 + 2 (2+\kappa) \epsilon$.}

It is interesting to give to the threshold value $g=1$ a more intuitive interpretation.
Recall that $g = Na/\lambda$, where $N$ is the number of strategic agents and $a$ 
the coefficient that relates the strength of the (apparent) signal to the investment 
volume. It is clear that the prediction of the future return must be compared to the volatility 
of the asset; therefore $a \sim v_0/\Sigma_0$, where $v_0$ is the average volume of 
investment for an individual agent. On the other hand, if the number of 
non strategic agents is $N_0$, one expects that the root mean square of $\Omega_t$ should scale 
like 
$\sqrt{N_0}$. Therefore $\Sigma_0 \sim \sqrt{N_0} v_0/\lambda$ 
(assuming that non strategic agents invest a similar volume $v_0$). Finally:
\be
g \sim \frac{N}{\sqrt{N_0}},
\ee
independently of both $v_0$ and $\lambda$. The conclusion is that the market enters the
`convention' phase as soon as $N > \sqrt{N_0}$. Hence 100 correlation-hunting traders
are enough the change qualitatively a market of 10000 non strategic agents.

\subsection{Overreaction to news} 
\label{causal}
Suppose now that there indeed exists a {\it small} objective correlation between 
$\delta P_{t+1}$
and $\delta I_t$, justified by some real economic mechanism relating the two quantities. This
means that the `noise' $\Omega_{t+1}$ which governs the price dynamics in the absence 
of strategic
traders and $\delta I_t$ have a non zero correlation coefficient:
\be
E[\frac{\Omega_{t+1}}{\lambda}\delta I_t]= \beta E[\delta I_t^2],
\ee
where we conform to  the common usage of calling this particular correlation coefficient the 
`beta'. The effect of such a term is to add a linear contribution to the effective potential 
$V(C)$
of the Langevin equation, which plays the r\^ole of a {\it symmetry breaking field} 
in the language of phase transitions \cite{Goldenfeld}:
\be
V(C) \longrightarrow V(C) - \beta C.
\ee
For $g < 1$ and $\beta$ small, the most probable value of $C$ is $\beta/(1-g)$  
($=\beta$ for $g=0$, as it should). Therefore, $C$ is of the same order as its
`true' cause whenever $g < 1$. However, in the limit $g \to 1^-$, the apparent 
correlation that arises becomes much larger than its true cause: the sensitivity of the market
to external information is anomalously amplified. For $g > 1$, the term 
$\beta C$ breaks the symmetry between the two conventions $\pm C^*$. In the limit
$\epsilon \to 0$, the most probable value of $C$ is given by $+C^*$ for 
$\beta \to 0^+$
and by $-C^*$ for $\beta \to 0^-$ (see Eq. \ref{Bolz}). Therefore, in the convention phase, 
the amplitude of the
apparent correlation is totally unrelated to that of the true correlations, although the {\it sign}
of the correlation reflects the underlying economic reality. One observes here a typical 
example of overreaction to news leading to excess correlations that are well documented in the literature
\cite{Thaler}. For example, the 
correlations between the stocks belonging to an index and the index itself are too strong 
to be explained by the intrinsic correlations between the stocks \cite{Schiller}. The present period (first 
quarter of 2003) is a good illustration of this effect: the cross correlations between U.S.
stocks is at a historical high; due to the large uncertainty, traders' hunt for useful 
information is more acute, and the influence of the index on individual stocks is 
expected to be anomalously large. In our model, this corresponds to the case where the indicator $I_t$ is the stock index, a case detailed in section \ref{s:index}. Another well known example is the
excess correlation (in particular in crisis periods) between emerging country markets belonging to different geographic regions \cite{Longin}. In order to understand the relation between these
effects and the present model, we need to make the following remark: although 
$C_t$ is a correlation between unequal times, the {\it equal time} correlation between 
$\delta P$ and $\delta I$ measured on a coarser time scale will reflect the value of the 
{\it lagged} correlation $C$. 
In other words, causal correlations on a fine time scale do generate equal time correlations 
on a coarser time scale. More precisely, one has:
\be
C^{(n)} \equiv E\left[(P_{t+n}-P_t)(I_{t+n}-I_t)\right] = 
E\left[\sum_{t'=t}^{t+n-1} \delta P_{t'} \sum_{t'=t}^{t+n-1} \delta I_{t'}
\right] = \sum_{t'=t+1}^{t+n-2} C_{t'}  \approx (n-1) C_t 
\ee
where the last equality holds if $n \Delta t \ll \tau$, i.e. when the coarse time increment $n
\Delta t$ is small
compared to the convention shift time $\tau$. Therefore, if strong causal correlations are
established intra-day, as is the case between individual stocks and the index, an excess daily 
correlation between stocks will also appear, see section \ref{s:index}.

\subsection{Consequences for the price fluctuations: excess volatility}

The feedback effect leads to an increase of the volatility of the price, since 
the instantaneous volatility is given by:
\be
\Sigma_t^2 = E[(\delta P_t)^2] = \Sigma_0^2 (1+ g^2 C_t^2) + O(C^4).
\ee
The non trivial dynamics of $C$ therefore leads to a volatility increase, which can
be substantial in the convention phase. This mechanism, interestingly, also leads to
to volatility fluctuations (or `heteroskedasticity'). These volatility fluctuations
are characterized by the correlation time $\tau$, which become large when $g$ 
approaches or exceeds the threshold value $g=1$ (see Eq. (\ref{Arrh})).

The above mechanism can easily be extended to the case where agents scrutinize 
$M$ different sources of information, say $I_t^k$, with $k=1,...M$. If the variation
of these `indices' are uncorrelated, it is easy to see that the simultaneous effect of all
the different feedbacks leads to a volatility given by:   
\be
\Sigma_t^2  = \Sigma_0^2 (1 + \sum_{k=1}^M g_k^2 C_{k,t}^2)+ O(C^4)
\ee
(with obvious notations). Therefore, the volatility can be substantially increased 
if a large number of information sources are overly interpreted. Within the context of 
efficient markets theory, all decisions are based on some `real' information. This 
corresponds, in the above formula, to choosing $\Sigma_0=0$ (the price movements are
all information-based) and $C^0_{k,t} = \beta_k$, where $\beta_k$ describes the `true' causal 
relation between an economic indicator $I^k$ and $P$. What we have shown here is that
because of the feedback loop, the empirical correlations (which are the only 
way to `learn' the value of the $\beta_k$'s in the absence of any firmly grounded theoretical 
model) are distorted and amplified, leading to a much larger apparent $C_{k,t} \gg C^0_{k,t}$, 
and therefore to a potentially considerable increase of the volatility as compared to 
its `fundamental' value.

We believe that the above scenario for self-referential speculation is generic. When  
strategies are built using the outcome of past random events, a feedback loop can appear and
destabilize the market from its putatively efficient behaviour. If the feedback is strong enough, a non trivial equilibrium can set in, 
where self-fulfilling prophecies can establish and survive. These conventions can
have no rational basis whatsoever, or be the result of the amplification of a very small, 
but indeed objective, correlation. 

\section{Price based strategies and market phases} 
\label{sect: autoreferential strategies}

\subsection{Motivations}
\label{micro}
As recalled in the introduction, the basic tenet of the theory of efficient market is that prices instantaneously 
reflect all useful information. However, since all market participants face impact and slippage issues
(due to the finite liquidity of any traded asset),
those who believe that they have some useful information about future price changes 
must use it in such a way that their very action does not perturb too much the market. 
Otherwise, the potential gain associated to this information cannot be realized, or only 
on small volumes. Therefore, informed investors must, to some extent, `dilute' their order in time. Doing so, 
they create positive temporal correlations -- the slow incorporation of
information into price is the `underreaction' phenomenon described in \cite{Shleifer,Stein}. 
Other participants that see an increase of price can believe that it is due to some 
information not yet available to them, but that is reflected by the recent price change. 
These participants will be tempted to `jump in the bandwagon' and act as trend 
followers: this is at the heart of the models developed in \cite{Shleifer,Stein}. Conversely, large orders in temporarily illiquid
markets might affect the price too much (`overreaction'), and some restoring trades 
will later on move the price back to a more realistic values.

Hence, there might indeed be deep reasons for which it can be useful to watch past price changes
and be influenced by them. That this is the case is practice is beyond any doubt, and is
confirmed by casual observation of traders in market rooms and by several formal 
surveys (\cite{Schiller}, p. 47). In fact, it seems that price itself is, for many
traders, the most relevant source of information (if not the only one, in the case of some
hedge funds using statistical methods). As in many previous models, we thus consider that the 
information used by some agents to predict future prices is the past price change itself. 
However, at variance with some of these models, the economic reality of the correlations is 
in fact not needed, since in the strong feedback phase these correlations may spontaneously 
appear.

\subsection{Trend following and contrarian conventions}

\begin{figure}
\begin{center}
\psfig{file=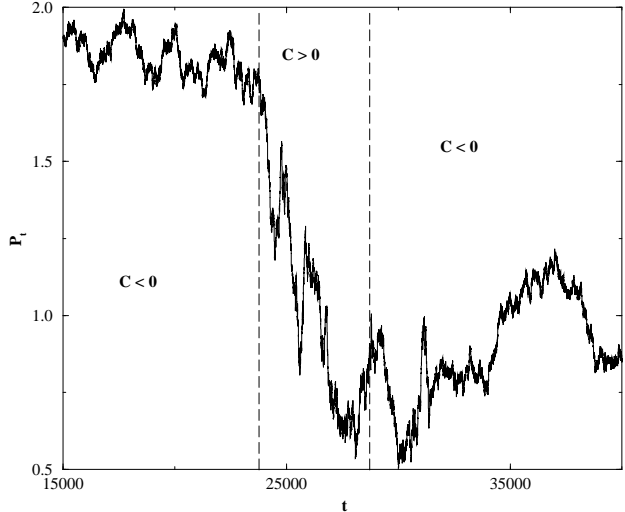,width=7cm,angle=270} 
\end{center}
\caption{Example of a synthetic price history, with two convention changes, for
$g=1.2$ and $\epsilon=0.01$. Note that the coarse grained volatility is smaller in 
contrarian phases ($C <0$) and larger in trend following phases ($C > 0$).}
\label{Fig3}
\end{figure}

In this section we therefore study the model where $\delta I_t = \delta P_t$.
In this case, the above correlation
coefficient $C_t$ becomes the autocorrelation of successive price changes. The above analysis
is almost unchanged, up to a renormalisation of the coefficient $h$ that appears in 
the non linear term $h C_t^3$. This comes from the fact that the denominator in the linear 
filter, namely $E[\delta P_t^2]$, is now itself affected by the feedback effect. Hence, the 
phase transition found above for $g=1$ is also present in this case. In the convention phase $g > 1$, the two states of the markets correspond 
to positive autocorrelations ($C = +C^*$), which can be called a {\it trend follower} phase where 
past price changes tend to be followed by a change of the same sign, or to negative 
autocorrelations ($C = -C^*$) in the {\it contrarian} phase, where 
past price changes tend to be followed by a change of opposite sign. Let us emphasize that 
a `trend following' period is not necessarily a period where the price steadily increases
(or decreases), but rather a period where successive price changes have a large probability
to be of the same sign (see the central period in Fig. \ref{Fig3}, corresponding to $C > 0$).

\begin{figure}
\begin{center}
\epsfig{file=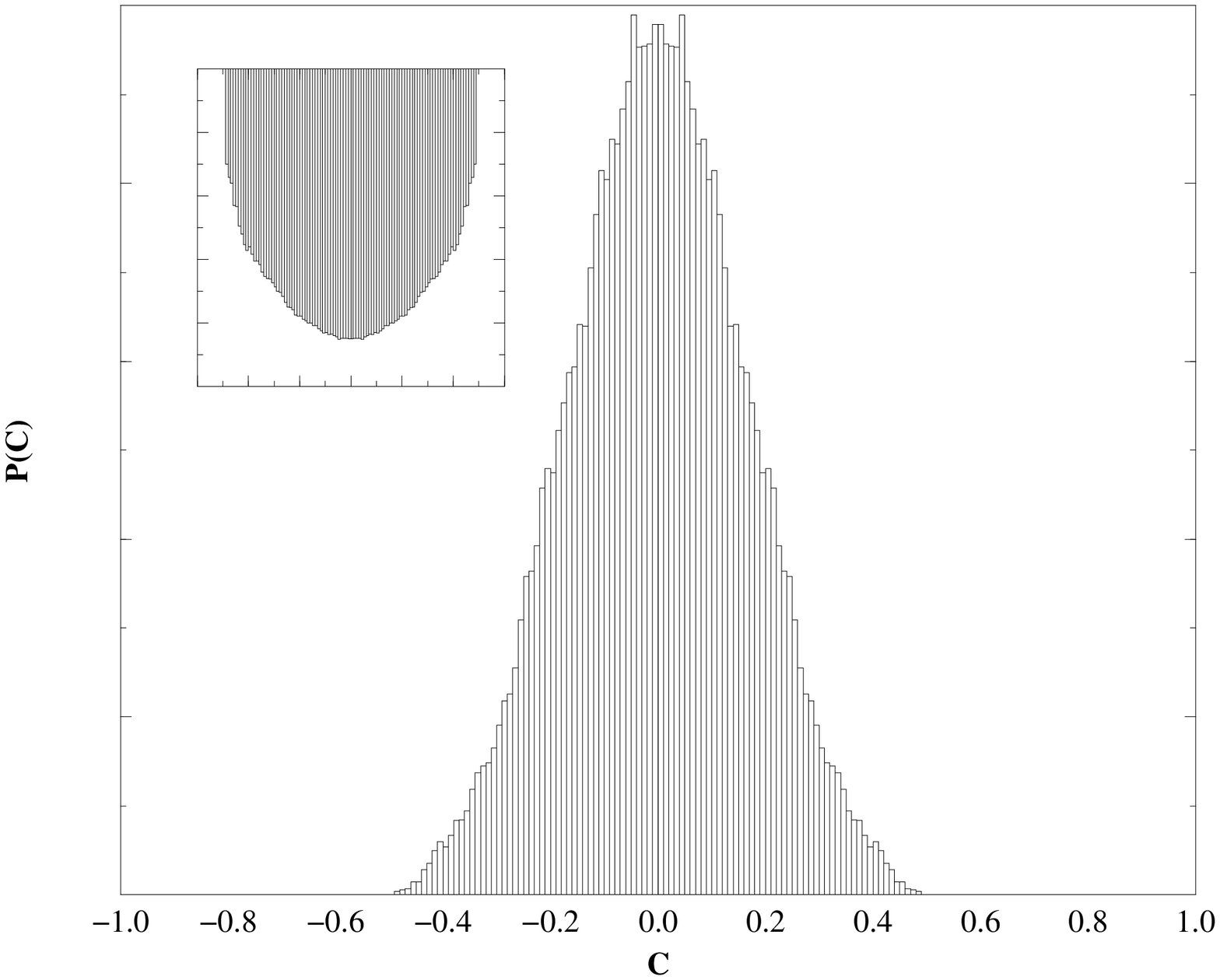,width=5cm,angle=270}\hspace{1cm}
\epsfig{file=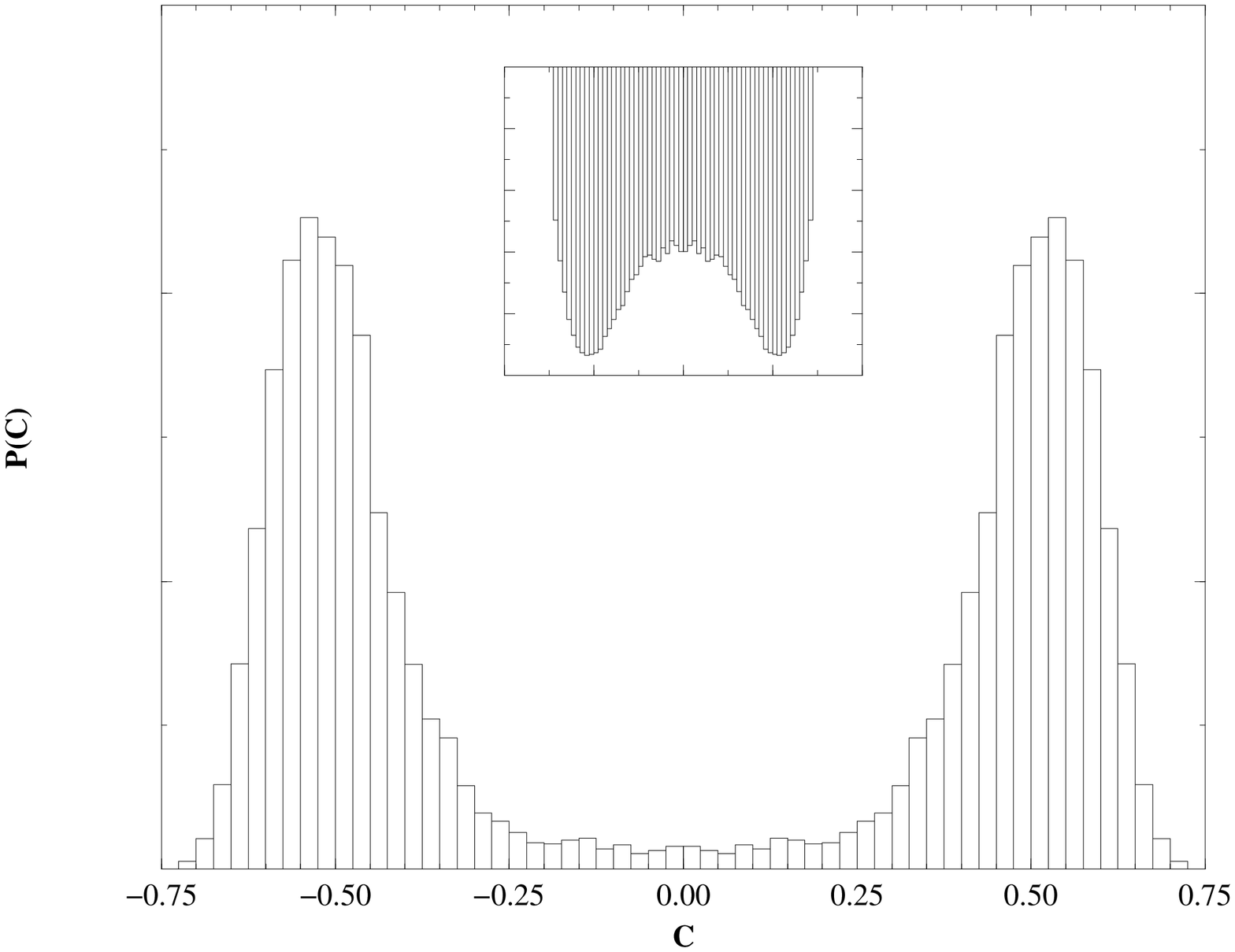,width=5cm,angle=270}
\end{center}
\caption{Left: Correlation histogram $P(C)$ for $g=0.9<1$ and $\epsilon=0.01$.
Right: Correlation histogram $P(C)$ for $g=1.3>1$ and $\epsilon=0.01$.  Insets: 
Effective potential $V(C)=-\ln P(C)$.}
\label{Fig2}
\end{figure}

We show in Fig. \ref{Fig2} 
the histogram of $C_t$ from the numerical simulation of Eq.(\ref{price}) with $\Omega_t$ a white Gaussian
noise and for two values of $g$. Note that without a
symmetry breaking term, the average autocorrelation is zero even for $g > 1$. We have here
an interesting statistical process where the long time autocorrelation is zero, but where
locally trends or anti-trends can appear and remain for quite long times. As mentioned in
the introduction, this corresponds to the market folklore: practitioners often talk about 
market phases where trend following strategies are supposed to be profitable, and market phases 
where contrarian (mean reverting) strategies are supposed to work. Interestingly, any long 
term analysis of the average correlation coefficient would fail to reveal such phases.

\subsection{Consequence on the volatility}

An important consequence of the existence of conventions is that the coarse-grained
volatility can be different from the instantaneous one. As above, the instantaneous volatility
is increased compared to its bare value $\Sigma_0$ and given by $\Sigma^2 = 
\Sigma_0^2(1 +g^2 C_t^2)$. On the other hand, the coarse-grained volatility $\Sigma_{cg}$, defined on an intermediate time scale 
$T^*$ such that $1 \ll T^* \ll \tau$ (such that $C_t$ itself has not evolved significantly),
is easily calculated to be:
\be\label{cgvol}
\Sigma_{cg,t} \approx \frac{\Sigma_0}{1-g C_t}.
\ee
This shows that the volatility is increased in the trend following convention and 
decreased in the contrarian convention. This is illustrated in Fig. \ref{Fig3}, where we show
the result of a simulation corresponding to $g=1.2$ and $\epsilon=0.01$, with a Gaussian noise term $\Omega_t$. 
Note that the true long time square volatility is equal to the time average of 
$\Sigma_{cg,t}$, and is dominated by the trend following phases.

Eq. (\ref{cgvol}) shows that the volatility can have large fluctuations, and long term
correlations; in particular, in the $g > 1$ phase, there are two time scales that govern
the evolution of $C_t$. One, relatively short one, governs the fluctuations of $C_t$ around the
dominant convention $\pm C^*$; the other, that can be much longer, is given by the 
flip time $\tau$ between the two conventions, Eq. (\ref{Arrh}). It might be tempting to relate this to the 
well known fact that empirical volatility fluctuations reveal non exponential, multiscale
relaxation in time \cite{volfluct1,volfluct2,Guillaume,Cizeau,PCB,Rama,LeBaron}.

\begin{figure}
\begin{center}
\epsfig{file=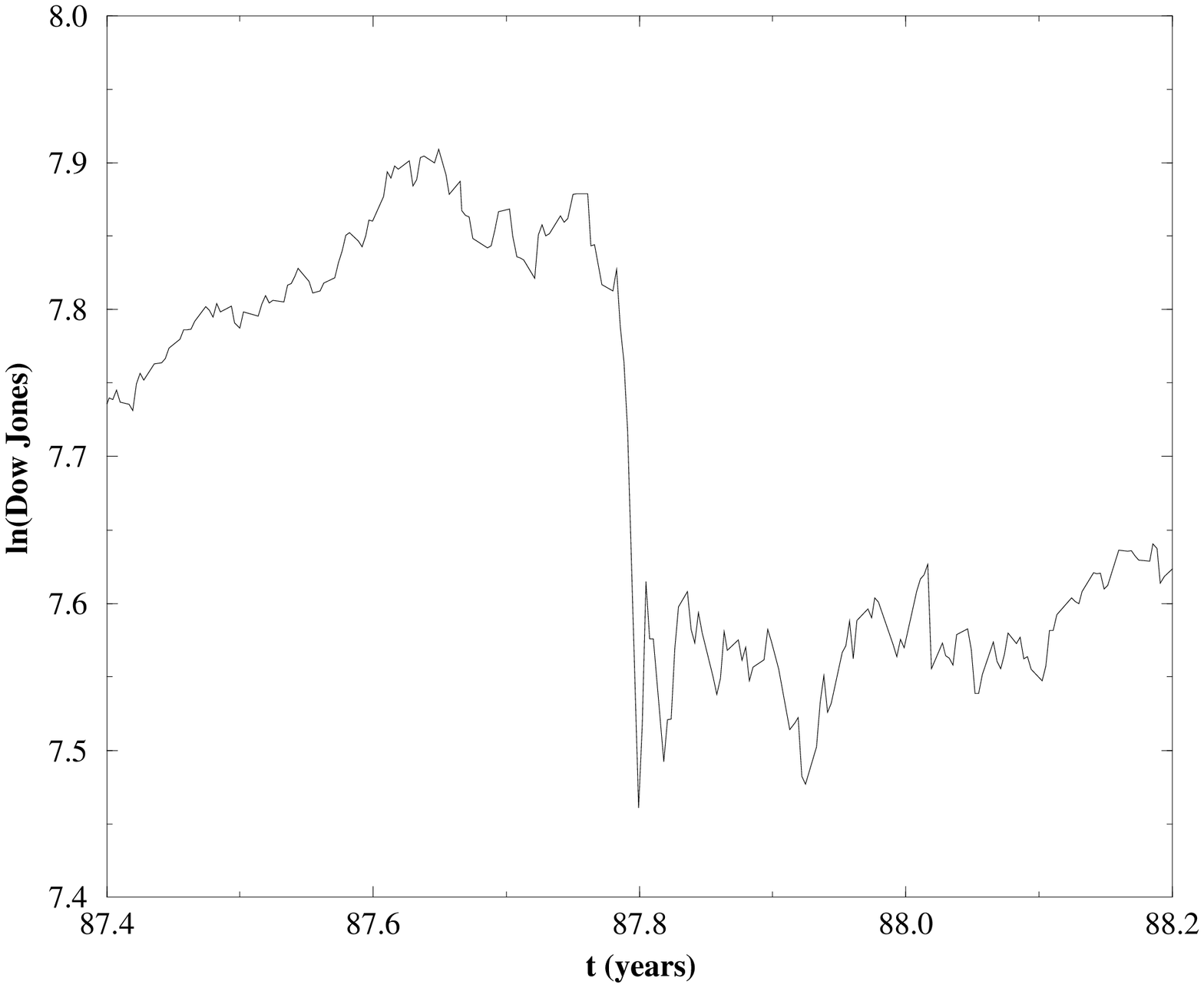,width=5cm,angle=270}\hspace{1cm}
\epsfig{file=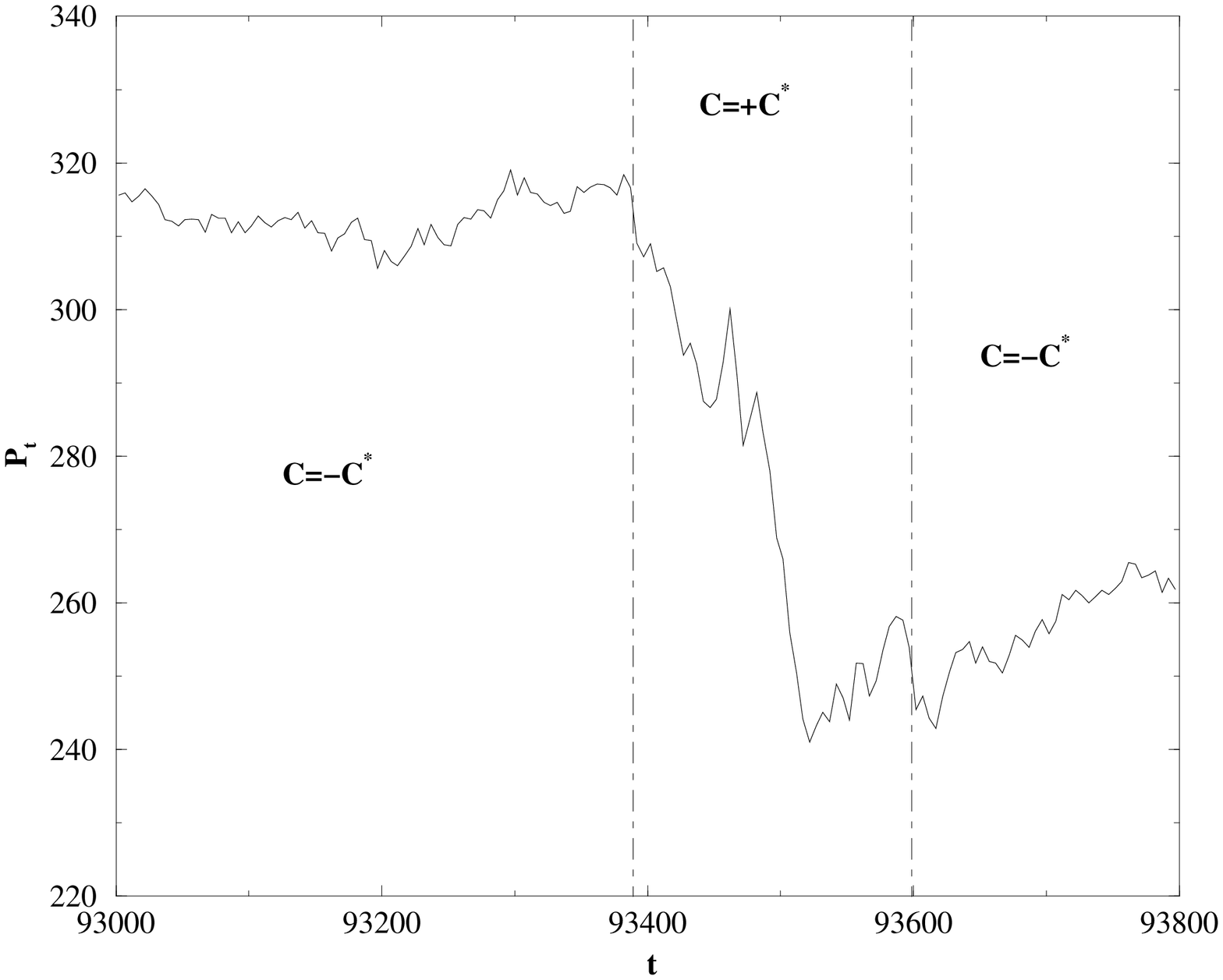,width=5cm,angle=270}
\end{center}
\caption{Left: Dow-Jones index (in logarithmic scale) around the 1987 crash. 
Right: Example of a sudden change of convention for $g=1.5$ and $\epsilon= 0.1$,
which mimics a crash. In order to `nucleate' the first convention 
change, the noise must accidentally imitate a trend following convention for a sufficiently 
long period of time (and vice versa for the second change).}
\label{Fig4}
\end{figure}

Suppose that one is in the trend following convention. The typical duration $\cal T$ 
of a trend can 
be obtained by comparing the value of the coarse grained volatility to the instantaneous one:
\be
{\cal T} \approx \frac{1 +g C^*}{1 - g C^*}, \qquad (gC^* \ll 1)
\ee
Hence, one can observe two types of dynamics when $g>1$. If the change of convention is 
faster than 
the typical duration of a trend, i.e. if $\tau \ll {\cal T}$, one obtains the dynamics shown 
in Figure \ref{Fig4}-b, where a period of low volatility is followed by a few sudden trends, 
which can be of any sign. Note that this can only occur if $\epsilon$ is large enough, which 
corresponds
to a very short memory time, in other words that agents over-focus on very recent events. 
The phenomenology is in this case quite different from that 
shown in Fig. \ref{Fig3}, where the price displays many trends before changing conventions. 

Up to now, we have implicitly assumed that the parameters $g$ and $\epsilon$ were constant. This
allowed us to extract the salient features of the model. 
In reality, however, these parameters should themselves be thought of as time dependent. Remember that $g$ is larger when more traders 
(or larger volumes for the same number of traders) use past prices to decide on their action.
It is clear that clear trending periods will increase the confidence in the trend following 
strategy and increase $g$. Although the full discussion of this extended model is beyond the 
scope of the present paper (see conclusion), we expect that reality might actually be
a mixture of convention phases when $g_t >1$ and floating conventions or random phases when 
$g_t < 1$. In this framework, crashes may be viewed as moments where both $g_t$ and $\epsilon$ 
become large: in panic situations, one expects the memory time to become shorter as agents think that
immediate information becomes crucial. This can create a strong temporary trend 
following 
convention that leads to a crash: see Fig. \ref{Fig4}. As the convention accidentally 
flips over to the contrarian one, 
the volatility falls sharply and so does the perceived uncertainty. 
One can expect $g_t$ to become small again: the crash is over. For purposes of illustration
and anecdotal evidence, we show in Fig. \ref{Fig4} a blow up of the 1987 crash period and 
the result of a simulation of our model.

\subsection{A special case: regressing on the index}
\label{s:index}
It is interesting to discuss the special case where the 
information $I_t$ is the stock index itself. It is clear that in practice, the evolution of
stock prices on short time scales is strongly affected by the index, which is an immediately 
available piece of relevant information for all market participants. Let us call $P_t^j$ ($j=1,...,M$) 
the price of the $j$-th 
stock belonging to the index. Then, assuming the index is computed as a equi-weight average
over all the stocks, one has:
\be\label{index}
\delta I_t \equiv \frac{1}{M} \sum_{j=1}^M \delta P_t^j.
\ee
One the other hand, the feedback effect of the index on the stock price can be written as:
\be
\delta P^j_{t+1} = \Omega^j_t+ g_j C_{j,t} \frac{\delta I_t}{\Sigma_I^2} + O(C_j^3),
\ee
where $\Omega^j_t$ results from the trading not based on the index, $C_{j,t}$ is the
empirical covariance between $\delta P^j_{t+1}$ and $\delta I_t$, and 
$\Sigma_I$ is the index volatility. Using Eq. (\ref{index}) 
one therefore finds, in the simple case where all $g_j$ are equal:
\be\label{index2}
\delta I_{t+1} = \frac{1}{M} \sum_{j=1}^M \Omega^j_t + g C_t \frac{\delta I_t}{\Sigma_I^2},
\ee
where $C_t = \sum_j C_{j,t}/M$ is the covariance between $\delta I_{t+1}$ and $\delta I_t$.

In the case of the index, it is reasonable to think that the feedback is a very high frequency one; therefore $\Delta t = 1$ probably corresponds here to a few minutes. Summing (\ref{index})
from $t$ to $t+n$ defines the return on a coarse-grained scale $\Delta I_t$. Assuming that 
$n \Delta t \ll \tau$, such that $C_t$ is approximately constant, leads to:
\be
\Delta I_t =  \frac{1}{M} \sum_{j=1}^M \hat \Omega^j_t + g \hat C_t \Delta I_t, 
\ee
which is valid when $n \gg 1$. In the above equation, $\hat \Omega^j$ is the aggregate of
the noise $\Omega^j$ over the time interval $[t,t+n]$, and $\hat C_t=C_t/\Sigma_I^2$. Finally,
\be
\Delta I_t = \frac{1}{M(1-g \hat C_t)} \sum_{j=1}^M \hat \Omega^j_t.
\ee
If the $\Omega^j_t$ were uncorrelated from one stock to the next, and in the absence of 
feedback, the index volatility would be very small compared to that of stocks (of order $1/\sqrt{M}$). Empirically, though, the U.S. stock index volatility is found to be as high as a third 
of the individual stock volatility. Of course, one expects that the $\Omega^j_t$ are 
somewhat correlated, reflecting a common sensitivity to news. However, the correlation
between stocks expected from fundamental analysis is insufficient to explain the observed 
correlation (and therefore the volatility of the index)\cite{Schiller}. The model presented here shows that
a high frequency positive feedback leads to an increase of the index volatility by a factor 
$1/(1-g \hat C_t)$, which can be large. This increase is actually larger than the increase
of the volatility of individual stocks induced by the above feedback (the factor is in that case $\sqrt{1+g^2 \hat C_t^2}$.

\section{empirical evidence} 
\label{sect: datas}

The aim of this section is present some empirical data that support our contention that 
some anomalous correlations exist in financial markets, with persistence times which can
be very long (ten years or so). We first present the case of the bond index vs. stock 
index correlation, which is interesting from the point of view of the present model 
because it might represent an empirical realization of the convention shift scenario
predicted above. We then turn to the analysis of the (daily) lagged autocorrelations 
of the Dow Jones index during the 20th century, which are clearly found to be significant, and time dependent (positive -- trend following, or negative -- contrarian). 

\begin{figure}
\begin{center}
\epsfig{file=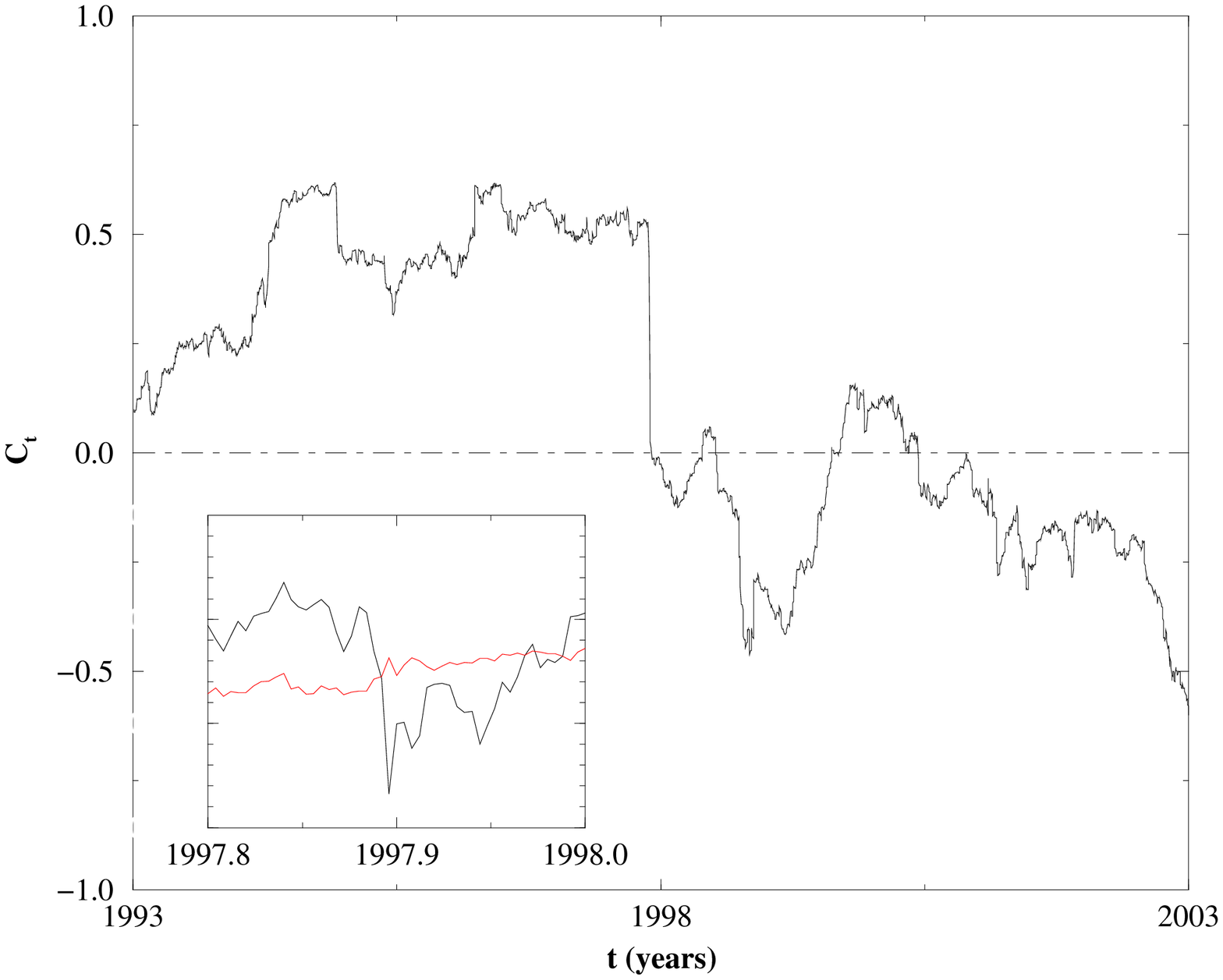,width=5cm,angle=270}\hspace{1cm}
\epsfig{file=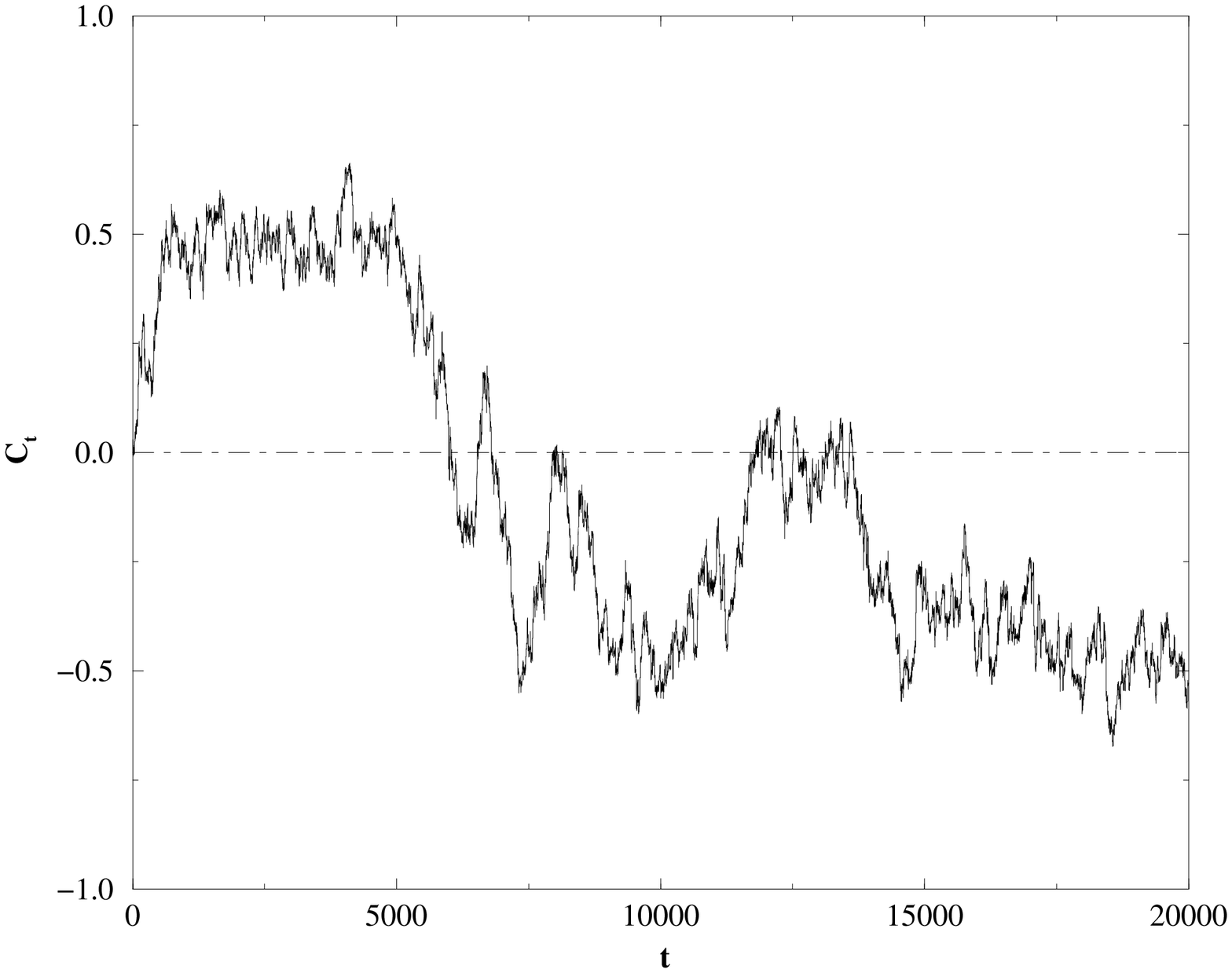,width=5cm,angle=270}
\end{center}
\caption{Left: Normalized correlation between the Dow-Jones daily returns and the daily returns of 
a U.S. bond index with 7 to 10 years bonds, computed with $\epsilon=0.01$. Note the convention change occurring at the end of 1997.
Inset: Evolution of the Dow-Jones and the bond index in the last quarter of 1997. 
Right: Time dependent correlation $C_t$ in our model, for $g=1.2$ and $\epsilon=0.01$.}
\label{Fig5}
\end{figure}

\subsection{The bond/stock cross correlation}

A very interesting example of rapid convention change has taken place in the 90's, and
concerns the correlation between stock markets and bond markets. The usual argument is 
that as long term rates fall, not only holding bonds becomes less profitable (bond prices 
rise) but also borrowing long term money becomes cheaper. Therefore stock markets become 
more attractive, and stock prices rise; this leads to a {\it positive} correlation between 
bond price changes and stock price changes. We compute the time dependent autocorrelation $C_t$ as an exponential moving average, as given in Eq. (\ref{Ct}), where $I_t$ is the bond index and 
$P_t$ is the log price of the Dow-Jones.
This correlation is indeed found to be positive, and
very strong ($\approx 0.5$), in the beginning of the nineties (see Fig. \ref{Fig5}). However, another story now
seems to be dominant: a fall in stock markets signals an increased anxiety of the operators
who sell their risky paper and buy non risky Government bonds. This has been called 
`Flight to Quality'. The result is a {\it negative} correlation between stock
prices and bond prices. Fig. \ref{Fig5} shows very clearly that a change of convention has taken
place in late 1997; the negative correlation is even stronger now (early 2003). Quite interestingly, this convention shift has taken place very abruptly due to a series of extreme events both on the stock market and 
on the bond market (see the inset of Fig \ref{Fig5}), as would predict the model discussed in this paper. [Note that we consider
here equal time correlations of daily returns; using the argument presented in section \ref{causal},
we expect a high frequency causal correlation to manifest itself as an equal time correlation
on a coarser time scale.]

\subsection{The Dow Jones}

We considered the detrended Dow-Jones index in the period 1900-2003, where the average return
was subtracted.
We have actually first fitted the log Dow-Jones as a second order polynomial in time, 
since the average return itself seems to have significantly increased between 1900 and 2000.
\footnote{Taking the raw returns without detrending in fact leads to the same conclusions.
The reason is that the typical daily returns are much larger than the average trend.}
We again compute the time dependent autocorrelation $C_t$ as an exponential moving average, 
as given in Eq. (\ref{Ct}), with now $\delta I_t=\delta P_t$, and where $P_t$ is the log price.
Since the returns are non Gaussian, we compared all our results with a null hypothesis 
benchmark where all returns are multiplied by random independent signs, such as to keep the 
correct statistics of the amplitudes but remove all serial correlations. (Note however that in 
this procedure, the correlation in the {\it volatility} is preserved.)

\begin{figure}
\begin{center}
\epsfig{file=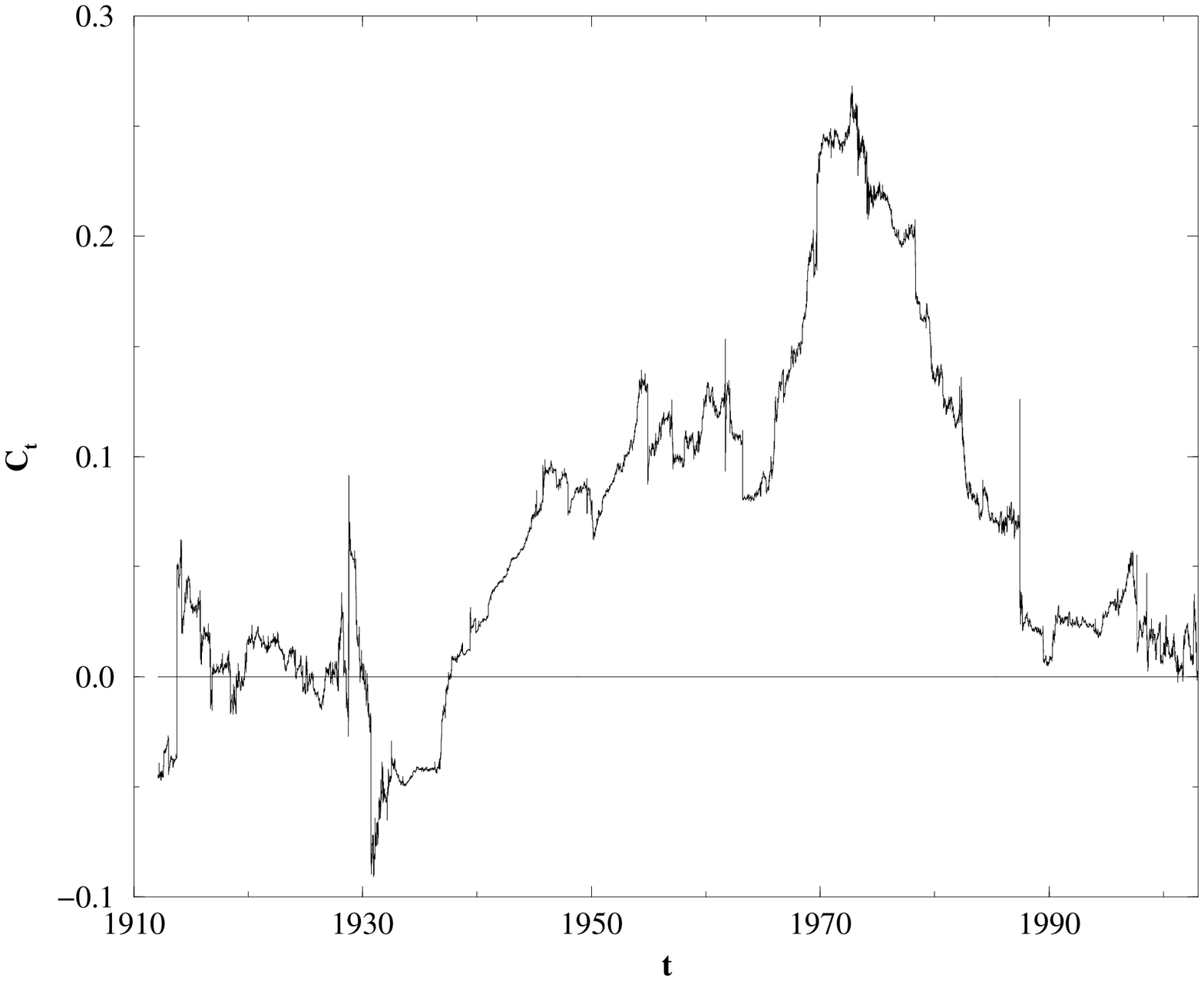,width=5cm,angle=270}\hspace{1cm}
\epsfig{file=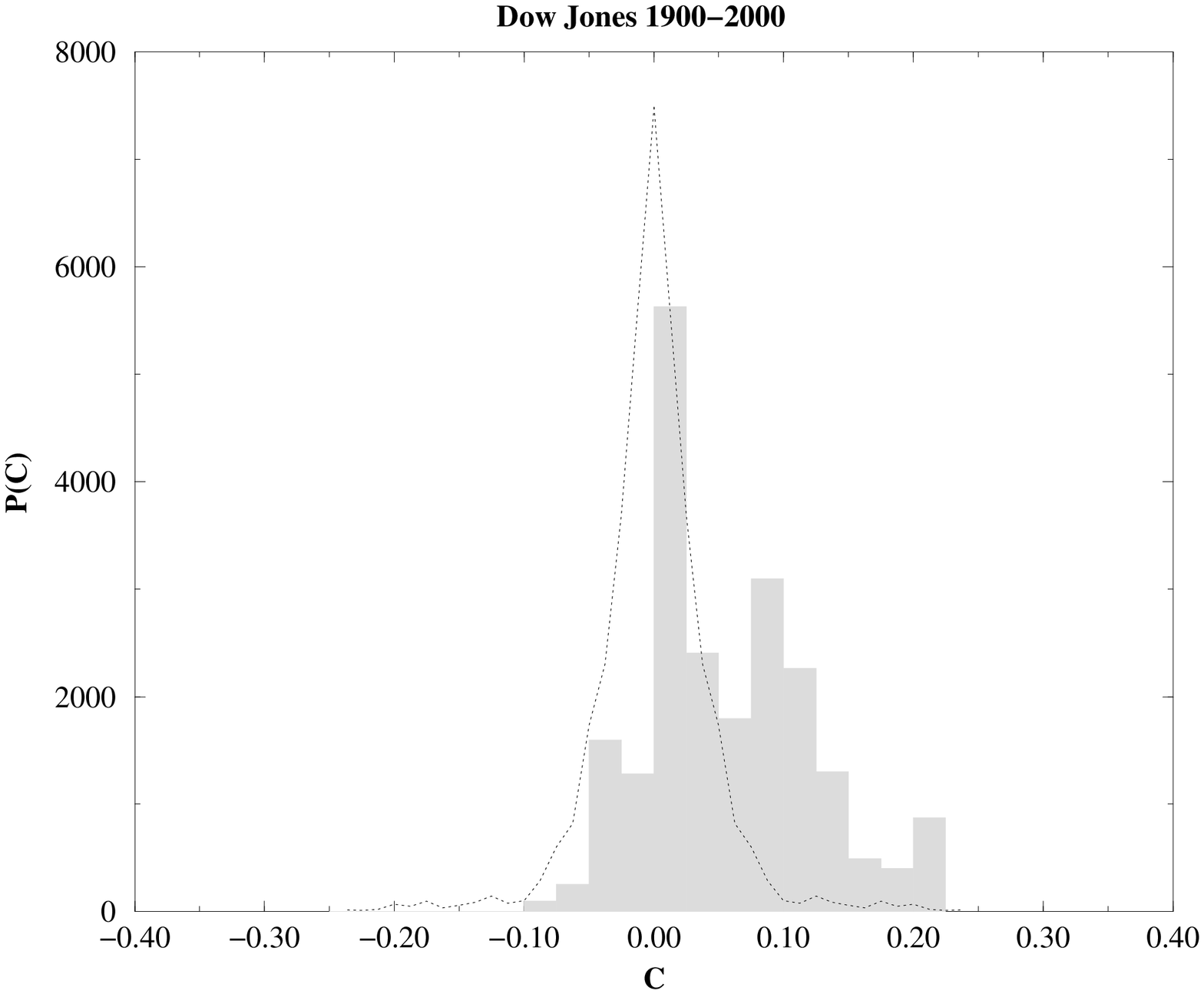,width=5cm,angle=270}
\end{center}
\caption{Left: Historical time series of the daily autocorrelation $C_t$ of
the Dow-Jones index, computed with $\epsilon=0.001$. Right: Correlation histogram $P(C)$ for the Dow-Jones with $\epsilon=0.001$, compared to 
the histogram computed with the same data and the same value of $\epsilon$ but with returns multiplied by random signs (dotted line).}
\label{Fig6}
\end{figure}

\begin{figure}
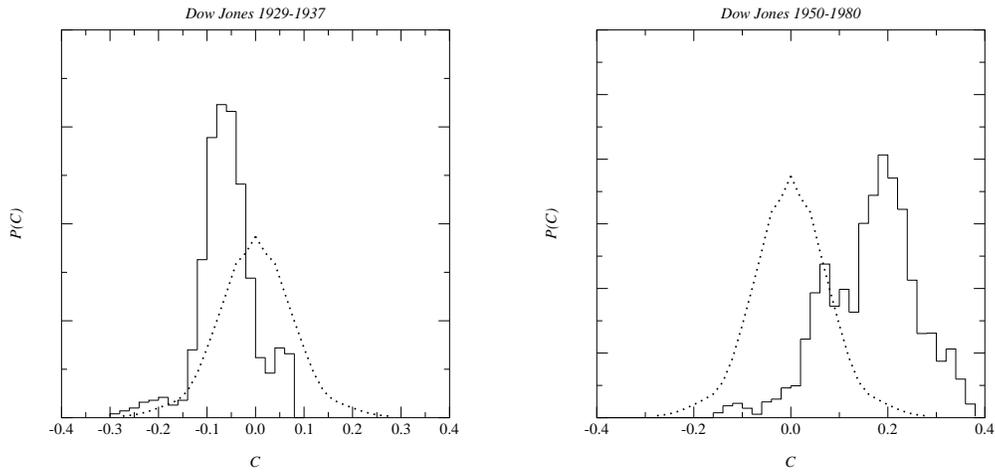

\begin{center}
\epsfig{file=after29.eps,width=6cm}\hspace{1cm}
\epsfig{file=after50.eps,width=6cm}
\end{center}
\caption{Left: Correlation histogram $P(C)$ for the Dow-Jones in the post-crash period 
1929-1937, with $\epsilon=0.01$, compared to the `zero correlation' histogram computed by
multiplying the returns by random signs (dotted line). Right: Correlation histogram $P(C)$ for the Dow-Jones in the trend following period 
1950-1980, with $\epsilon=0.01$, again compared to `zero correlation' histogram (dotted line).}
\label{Fig7}
\end{figure}

We show in Fig. \ref{Fig6} the time evolution of $C_t$ computed for $\epsilon=0.001$  
for the real time series. One clearly sees that (i) $C_t$ can be substantially larger than expected 
if no correlations were present and (ii) the time scale for the evolution of $C_t$ can be much 
larger than $1/\epsilon \approx 3$ years. 
Plateaus that last several decades can be observed. The histogram of different values of 
$C_t$ is shown in Fig. \ref{Fig5} and is markedly different from the one corresponding `scrambled' series, for which all correlations are killed. (The hypothesis that the
two distributions are the same is strongly rejected by the Kolmogorov-Smirnov test). The century was dominated by 
a positive correlation convention, especially between the 50's and the 80's. 
Nevertheless the negative correlation convention seems to appear after the 1929 during the Great 
Depression. There are also regimes where $C_t$ is close to zero. This suggests that in fact $g$ has 
varied over the years, with periods where $g < 1$, with no clear trends nor anti-trends appearing, and
periods where $g > 1$, during which the market is `locked' in one convention or the other.
In order to check whether the plateau values appearing in Fig. \ref{Fig5} do indeed correspond to conventions, 
we have determined to probability distribution of $C_t$ with a smaller averaging time 
of 100 days ($\epsilon=0.01$), and in restricted periods of time: (a) at the beginning of the 30's 
(contrarian convention) and (b) between the 1950 and 1980 (trend following convention), see 
Fig. \ref{Fig7}. The comparison with `scrambled' data indeed shows a clear assymetry in both cases, that should not exist if all serial correlations were zero. 

These curves show that conventions can persist up to 30 years. The change in 
convention can be rather smooth, like during the second part of the century. As we saw before, 
the value of the most probable value $C^*$ is related to $g$, i.e. to the number of 
agents using a self-referential strategy. Then these smooth changes 
can be explained by a continuous change in the number of these agents. 
A change of convention can also occur suddenly, triggered by an extreme event, like it did after 
1929. It can be explained as suggested above: before the crash, $g$ is smaller than unity and no clear 
convention exists. The crash induces an enormous uncertainty about the true value of 
stocks, and encourages agents to pay more attention to past price variations. This may have led
to a substantial increase of $g$ that favored the appearance of a contrarian convention. 

From the data, it appears that there might be a systematic bias towards the trend following convention. One in fact expects that some symmetry breaking term, favoring $C > 0$, should exist in general: first, as
mentioned in section \ref{micro}, there might be good reasons to think that positive correlations are indeed
created by the time dilution of large orders, or other mechanisms. Also, for purely psychological reasons,
trend following strategies are more likely to be adopted than contrarian strategies, because the pattern is much more obvious. This 
can be modeled by postulating that $g$ depends on the sign of $C_t$, with $g_+ > g_-$. Therefore, 
nicely symmetric histograms such as those presented in Fig. \ref{Fig2} are unlikely to be observed in real markets.

\section{Conclusion and Perspectives} 
\label{sect: conclusion}

In this paper, we have defined and studied a generic, parsimonious model that describes the 
feedback effect of self-referential behaviour. If sufficiently strong, this feedback 
destabilizes the market and non trivial correlations can spontaneously appear, or be 
anomalously 
amplified. In this case the market enters a new equilibrium state where strong correlations 
between
a priori uncorrelated quantities might self consistently establish. These anomalous 
correlations
lead to both excess volatility (that may display long memory) and excess cross correlations.
An interesting outcome of our model is (i) the existence of correlations with an amplitude 
unrelated with any `rational' value, and (ii) the appearance of a very long switching time scale,
unrelated with the natural time scales of the dynamics (i.e. the `decision' time scale
$\Delta t$ or the memory time $T$). Therefore, our model displays {\it regime switching} 
over long times scales, which by the way justifies why agents should use a finite memory 
time in order to measure correlations in an ever changing environment. 
It is also worthwhile emphasizing that if the price itself 
is used as a source of information, some 
linear autocorrelations do appear on intermediate time scales, but average out on time scales
larger than this switching time. In other words, the price process has zero average 
linear correlations, but non zero {\it local} autocorrelations (trending or mean reverting).
We have presented convincing empirical evidence that such conventions exist in financial 
markets; one of the most compelling case concerns the correlation between stock markets 
and bond markets, where both market `states' can be observed: the correlation appears to have rapidly 
shifted in the last decade from being strongly positive to being negative.

The model could be extended in several directions. First, one could consider serial 
correlations beyond the elementary time lag $\Delta t$, say between increments lagged 
by $n \Delta t$. At the linear level, the stability criteria is easily shown to be $g_n < 1$,
where $g_n$ is the feedback strength corresponding to lag $n \Delta t$. However, interesting 
non linear effects can appear. For example, with two lags $n=1$ and $n=2$, one can observe 
a `first order' phase transition where the most probable value of $C$ discontinuously jumps from $C=0$ to $\pm C^*$, with $C^* > 0$ even close to the transition. This is distinct from the 
`second order' scenario explored in the present paper, where $C^* \sim \sqrt{g-1}$.
Second, one could consider the case where the different sources of informations 
$\delta I_t^k$ are themselves asset prices is quite interesting since the feedback loop also affects the
cross-correlation between the different assets. Non trivial coupled convention dynamics
can set in, in particular when the number of assets is small. 

An ingredient that should be implemented is the feedback between the values of the coupling 
parameter $g$ and memory time $1/\epsilon$, and the past price dynamics itself. As we have seen above, the value of $g$ 
is related to the number of agents (or more precisely the total volume of orders) 
that act in a self-referential way. It is clear that both in periods of large uncertainty 
(after a crash, for example) or within a speculative bubble where the trend following
strategy appears to be successful, one expects the value of $g_t$ to grow. (A similar 
mechanism was recently considered within the `Grand Canonical' version of the Minority Game, see \cite{GCMG,GB2,GB})
It would be interesting to study a precise model where the dynamics of $g_t$ and 
that of the price and volatility are explicitly coupled, in the spirit of \cite{Langevin}. 
One can expect that such a model would be able to capture a lot of the financial markets 
phenomenology. Along similar lines, if there are several sources of information $\delta I_t^k$,
one should expect that the feedback of successful strategies onto the value of the
couplings $g_{k,t}$ will be unstable in the sense that one of the $g_k$ will grow at the 
expense of the others, because the coordination of strategies leads to stronger self-fulfilling
prophecies and therefore larger potential profits. In other words, this feedback between the
tendency to follow a pattern and its predictability leads to a condensation of the strategies 
in a few prominent conventions, with abrupt transitions between those. 

Finally, we need to discuss the above model from the point of view of Efficient Markets, and 
show how it should be modified to describe the {\it long term} behaviour of market prices.
If the information is systematically over-interpreted and the volatility much too large compared 
to that of the `fundamental' value, the price should go on long time 
scales to completely unrealistic values. More precisely, the difference between the `fair' price and 
the market price would typically grow as $\sqrt{T}$. [In mathematical terms, the market price 
and the fundamental price would not `cointegrate'.]. The answer to that paradox is that in fact 
nobody knows the fair price of a stock more accurately than within, say, a factor of two. This was 
actually proposed by Black \cite{Black} (somewhat humoristically) as the definition of an 
efficient market, and this view seems to us to be fundamentally correct. For example, 
the historical analysis 
presented in \cite{Schiller}, p. 8, shows that the price to earning ratio of U.S. stocks has 
indeed fluctuated, from 1900 to 2000, between $10$ and $40$. So it is reasonable to think that 
there is a wide band of
prices across which arbitrage cannot take place, because of the lack of a reliable estimate
of what the true price should be. As emphasized by Shleifer and others \cite{Shleifer}, arbitrage
only makes sense if one can compare the {\it relative} price of two assets, but becomes 
very dodgy if one speaks about absolute values. Therefore, one expects that as long as
the price is within a factor two of the `true' price, no mean reversion term, induced
by the presence of arbitrageurs, needs to be added to our dynamical equation for the price, 
Eq. (\ref{price}). Mathematically, this mean reversion effect is described by adding to the right 
hand side of Eq. (\ref{price}) a term proportional to $-\kappa \log (P_t/P_0)$, where $\kappa$ 
measures
the strength of the demand driven by fundamental considerations, and $P_0$ the true fair price 
(see also \cite{Langevin}, where this term was introduced). On short time scales, or if $\kappa$ 
is sufficiently small, this term 
can be neglected and the analysis presented above should be valid. On long time scales, however, 
such that the random fluctuations become of the order of (say) 100 \%, one should expect these mean 
reversion effects to become relevant. For a
typical stock with a daily volatility of $3 \%$, this corresponds to 1000 days, or four years.
Such a time scale is precisely the typical reversion time scale discovered by de Bondt and
Thaler in their paper on overreaction in stock markets \cite{Thaler}. Hence, in a world where absolute 
references are
lacking, one expects that the short to medium time scale dynamics of markets will be 
dominated by 
the self-referential effects described in the present paper.

\section*{Acknowledgments}
We thank M. Potters, J.P. Aguilar, I. Giardina, L. Laloux and A. Matacz for some important remarks 
and help with the data. We also thank X. Gabaix and A. Orlean for very useful comments that 
helped improving the manuscript.

\end{document}